\documentclass[a4paper,longbibliography,superscriptaddress,twocolumn,showkeys,pre, aps ]{revtex4-1}
\usepackage{amsfonts,amsmath,amssymb,hyperref}
\usepackage[pdftex]{graphicx}
\usepackage{color}
\usepackage{times}
\usepackage{tikz}
\usepackage{epsfig}
 \usepackage{ulem}

\begin{document}

\title{Simulating non-Markovian stochastic processes}

\author{Marian Bogu\~n\'a} 
\affiliation{Departament de F\'{\i}sica Fonamental, Universitat de
  Barcelona, Mart\'{\i} i Franqu\`{e}s 1, 08028 Barcelona, Spain}
\author{Luis F. Lafuerza}
\affiliation{Theoretical Physics Division, School of Physics and Astronomy, University of Manchester, Manchester M13 9PL, UK}
\author{Ra\'ul Toral}
\affiliation{IFISC (Instituto de F\'isica Interdisciplinar y Sistemas Complejos), Universitat de les Illes Balears-CSIC, Palma de Mallorca, Spain}

\author{M. \'Angeles Serrano}

\affiliation{Departament de F\'{\i}sica Fonamental, Universitat de
  Barcelona, Mart\'{\i} i Franqu\`{e}s 1, 08028 Barcelona, Spain}

\begin{abstract}
We present a simple and general framework to simulate statistically correct realizations of a system of non-Markovian discrete stochastic processes. We give the exact analytical solution and a practical an efficient algorithm alike the Gillespie algorithm for Markovian processes, with the difference that now the occurrence rates of the events depend on the time elapsed since the event last took place. We use our non-Markovian generalized Gillespie stochastic simulation methodology to investigate the effects of non-exponential inter-event time distributions in the susceptible-infected-susceptible model of epidemic spreading. Strikingly, our results unveil the drastic effects that very subtle differences in the modeling of non-Markovian processes have on the global behavior of complex systems, with important implications for their understanding and prediction.  We also assess our generalized Gillespie algorithm on a system of biochemical reactions with time delays. As compared to other existing methods, we find that the generalized Gillespie algorithm is the most general as it can be implemented very easily in cases, like for delays coupled to the evolution of the system, where other algorithms do not work or need adapted versions, less efficient in computational terms.

\end{abstract}

\date{\today}
\maketitle

\section{Introduction}

Discrete stochastic processes are widespread in nature and human-made systems. Chemical reactions and biochemical processes in living cells, epidemic propagation in populations, and diffusion of information in societies and technological networks are all examples of systems whose states change at discrete random intervals, defining a sequence of events that conform a mixture of temporal point processes. In general, these processes are assumed to be memoryless, with future occurrences predictable based solely on the present state of the system, and with exponentially distributed inter-event times so that the dynamics can be described only in terms of the rates of occurrence of each of the processes involved. In this case, there exists stochastic simulation algorithms able to generate statistically exact realizations of the stochastic process, including the seminal method developed by Gillespie  for Markovian dynamics modeled by Poisson point processes and its variations~\cite{GILLESPIE:1976rw,GILLESPIE:1977db,Gillespie:2001kl}.

While considering that the time between two consecutive events is exponentially distributed turns out to be a good approximation in some cases, it fails dramatically in many others. Indeed, non-exponential inter-event time distributions have been reported in different contexts, making evident memory effects and, thus, non-Markovian dynamics. Applications include different problems in reliability analysis~\cite{Ushakov:2012} or queuing theory~\cite{Kella:2006}, but relevant examples are found in many different areas dealing, for instance, with patterns of human activity~\cite{Eckmann05102004,Barabasi:2005la,Malmgren:2008fk,Rybski04082009,Malmgren25092009,Stehle:2010uq,Zhao:2011fk,Fernandez-Gracia:2011fk} --from communication to mobility--, bursty fluctuations of connections in temporal networks~\cite{Salathe:2010vl,Gautreau02062009}, or biochemical reactions with time delays~\cite{Jensen:2003,Bratsun:2005fk,Barrio:2006uq,Roussel:2006sf,Cai:2007th,Anderson:2007ad,Brett:2013kx,Miekisz:2011,Lafuerza:2011b,Lafuerza:2011,Lafuerza:2013}, to name just a few. Non-Markovian stochastic processes are notoriously difficult to tackle analytically and, in many cases, their understanding relies on numerical simulations~\cite{Bratsun:2005fk,Barrio:2006uq,Roussel:2006sf,Cai:2007th,Anderson:2007ad,Fernandez-Gracia:2011fk,Van-Mieghem:2013db}. 

In this paper, we develop a simple and general framework to simulate statistically correct realizations of discrete stochastic processes, each with an arbitrary inter-event time distribution, that may stochastically create or annihilate other processes  and that can depend on the current state of the system. We provide the exact solution to the problem, along with an approximation in the limit of large systems leading to an efficient and simple stochastic simulation algorithm in the same spirit of the Gillespie algorithm for practical applicability. We apply our generalized Gillespie algorithm to two case studies: the susceptible-infected-susceptible epidemic spreading model in contact networks, and a system of coupled biochemical reactions with time delays where we compare with already existing methods. Our results highlight the important effects that subtle differences in the non-Markovian dynamical rules underlying the stochastic processes have on the global behavior of complex systems.

\section{Markovian stochastic simulation: The Gillespie algorithm}
The Gillespie algorithm~\cite{GILLESPIE:1976rw,GILLESPIE:1977db} was originally designed to simulate systems of coupled (bio)chemical reactions within a thermal bath in a well mixed environment but, more generally, it can be applied to any system of discrete Markovian stochastic processes. The algorithm takes advantage of the theory of superposition of a (fixed) number of renewal processes~\cite{coxbook:1965}. Suppose we have a collection of $N$ statistically independent discrete stochastic processes, each occurring at rate $\lambda_i$, $i=1,\cdots,N$. The Gillespie algorithm generates a sequence of events by specifying, at each step of the simulation, the time until the next event $\tau$ (generated from the distribution $\varphi(\tau)$) and the next event $i$, generated from the probability $\Pi(i)$. It can be proved that for Poisson point processes (constant rates) these are given by
\begin{equation}
\Pi(i)=\frac{\lambda_i}{N \bar{\lambda}} \mbox{ \hspace{0.2cm} and  \hspace{0.2cm}} \varphi(\tau)=N \bar{\lambda} e^{-N \bar{\lambda} \tau}, 
\label{prob_i_gillespie}
\end{equation}
where $\bar{\lambda}=N^{-1}\sum_{k=1}^N \lambda_k$ is the population mean rate of the set of processes~\cite{coxbook:1965}. Notice that, in general, the occurrence of a particular event can modify, besides the state of the system, both the rates and/or the number  of ``active'' processes (those that can occur given the current state of the system) for the next iteration. For instance, a given reaction taking place modifies the number of molecules of all species involved in that particular reaction which, in turn, modifies the rates of occurrences of all reactions in which these species participate. This makes the algorithm extremely powerful and versatile, as it can simulate reaction-like processes for which the number of processes is stochastically generated by the realization, including non-equilibrium dynamics with absorbing states. 

\section{Non-Markovian stochastic simulation}

Next, we generalize the Gillespie algorithm to account for non-Markovian inter-event times. As before, we consider a set of $N$ statistically independent discrete stochastic processes, each with an inter-event time distribution $\psi_i(\tau)$; $i=1,\cdots,N$~\cite{COX:1954fk}. Suppose now that, for a given process $i$ and a given point in time, we know the time elapsed since the last event, $t_i$, and ask the probability that next event will occur a time between $\tau$ and $\tau +d\tau$ from that moment (hereafter, we use latin symbol $t_i$ to denote elapsed times and greek symbol $\tau$ to denote the time until a future event). This probability density can be expressed as~\cite{cox:1970}
\begin{equation}
\psi_i(\tau |t_i)=\frac{\psi_i(\tau+t_i)}{\Psi_i(t_i)},
\label{eq:2}
\end{equation}
where $\Psi_i(\tau)$ is the survival probability of process $i$, that is, the probability that the time until the next event is longer than $\tau$, $\Psi_i(\tau)=\int_{\tau}^{\infty} \psi_i(s)ds$. Analogously, the conditional survival probability of process $i$ is given by $\Psi_i(\tau |t_i)=\int_{\tau}^{\infty}ds\,\psi_i(s|t_i)=\Psi_i(\tau+t_i)/\Psi_i(t_i)$.

In a single realization of the dynamics, all processes happen in the same timeline in a random order. Therefore, to generate a statistically correct sequence of events in a simulation, we have to evaluate the joint probability $\varphi(\tau,i|\{ t_k \})$ that, given the times $\{t_k\}$ elapsed since the last occurrence of each process up to a given point in time $t$, next event taking place corresponds to process $i$ and will occur at time $t+\tau$. Since the probability that process $k\ne i$ does not occur is $\Psi_k(\tau|t_k)$, we have
\begin{equation}
\varphi(\tau,i|\{ t_k \})=\psi_i(\tau |t_i) \prod_{k \ne i} \Psi_k(\tau |t_k)=\frac{\psi_i(\tau+t_i)}{\Psi_i(\tau+t_i)} \Phi(\tau|\{t_k\}),
\label{eq:3}
\end{equation}
where  
\begin{equation}
\Phi(\tau|\{t_k\}) = \prod_{k=1}^N\frac{\Psi_k(\tau+t_k)}{\Psi_k(t_k)}
\label{cumulative_tau}
\end{equation}
is the survival probability of $\tau$, i.e. the probability that no reaction occurs before $t+\tau$.
Note that the joint probability Eq.~(\ref{eq:3}) is well normalized. 

 Given the ocurrence time $\tau$, the probability that next occurring event belongs to process $i$ is
\begin{equation}
\Pi(i|\tau,\{t_k\})=\frac{\varphi(\tau,i|\{t_k\})}{\sum_j\varphi(\tau,j|\{t_k\})}=\frac{\lambda_i(t_i+\tau)}{\sum_j\lambda_j(t_j+\tau)},
\label{prob_i}
\end{equation}
where we have introduced the instantaneous (hazard) rate of process $k$ as
\begin{equation}
\lambda_i(\tau) \equiv \frac{\psi_i(\tau)}{\Psi_i(\tau)}.
\label{instantaneous_rate}
\end{equation}

Equations (\ref{cumulative_tau}) and (\ref{prob_i}) provide us with an algorithm that generates statistically correct sequences of events. Specifically:
\begin{enumerate}
\item
Initialize elapsed times for all processes.
\item
Draw a random time from the cumulative distribution Eq.~(\ref{cumulative_tau}), by solving $\Phi(\tau|\{t_k\})=u$, being $u$ a uniform random number in the interval $(0,1)$ and update current time as $t\rightarrow t+\tau$. 
\item
Choose a process $i$ from the discrete distribution Eq.~(\ref{prob_i}).
\item
Update the list of elapsed times as 
\[
t_k\rightarrow t_k+\tau, \, \forall k \ne i 
\mbox{ \hspace{0.5cm} and  \hspace{0.5cm}}
t_i=0.
\]
\item
update the state of the system and, if needed, the set of active processes. If a new process, say process $k$, is activated, set its elapsed time. Go to step 2. 
\end{enumerate}

The initialization of elapsed times can be implemented in different ways depending on the particular application. One simple possibility would be to set initial elapsed times to zero. Another approach is to assume that the system is already in the steady state and, thus, set elapsed times according to the probability density $\Psi_i(t)/\langle \tau_i \rangle$, where $\langle \tau_i \rangle$ is the average inter-event time of process $i$~\footnote{Suppose we pick any point in the time axis. For any renewal process, this point will always fall between two events. Since we do not have any other information, all what we can say is that the time interval between this two events is longer than the time from the chosen point until the next event. Therefore, the probability density of the time until the next event is proportional to the survival probability of the renewal process~\cite{cox:1970}}.

\subsection{Generalized Gillespie algorithm}

The most frequent applications typically involve a fairly large number of processes $N$. It is possible to work out a simple approximation that becomes exact in the limit $N\rightarrow \infty$ and drastically simplifies the numerical computation of the time $\tau$ needed in point 2 of the algorithm. We start by rewriting function $\Phi(\tau|\{t_k\})$ as
\begin{equation}
\Phi(\tau|\{t_k\})=\exp{\left[-\sum_{k=1}^N \ln\left( \frac{\Psi_k(t_k)}{\Psi_k(\tau+t_k)}\right) \right]}.
\label{function_phi}
\end{equation}
The sum within the exponential function is a sum of $N$ monotonously increasing functions of $\tau$. Therefore, when $N\gg 1$  the survival probability $\Phi(\tau|\{t_k\})$ is close to zero everywhere but when $\tau \sim 0$. Hence we only need to consider $\Phi(\tau|\{t_k\})$ around $\tau=0$, where an expansion in small $\tau$ can be performed: $\Psi_k(\tau+t_k) = \Psi_k(t_k)-\psi_k(t_k)\tau+\mathcal{O}(\tau^2)$.
Plugging this expression into Eq.~(\ref{function_phi}), we can write
\begin{equation}
\Phi(\tau|\{t_k\}) \approx e^{ -\tau N \bar{\lambda}(\{t_k\})},
\label{function_phi_approx}
\end{equation}
where the average rate is $\bar{\lambda}(\{t_k\})= N^{-1}\sum_{k=1}^N \lambda_k(t_k)$.
The previous expansion assumes that $\Psi_k(\tau+t_k)$ is analytical at $t_k=0$, a hypothesis which sometimes is not true. To overcome this singular case, we remove the last event, the one for which $t_\textrm{last}=0$ from the sum in $\bar{\lambda}(\{t_k\})$. This implies that the probability to choose the same event two times in a row with our algorithm is zero. While this restriction is in general not present in the real dynamics, the probability of such event is negligible for large $N$ and, thus, our assumption does not have any noticeable effect while avoiding a potential divergence of the algorithm in cases where $\lim_{\tau \rightarrow 0^+}\psi_i(\tau) = \infty$~\footnote{This, however, can be implemented exactly by generating two different times $\tau_1$ and $\tau_2$, the first for the process with diverging rate from its inter-event time distribution and the other from the rest of the processes, and then taking the minimum of the two.}.

Within this approximation, the probability that the next event taking place belongs to process $i$ 
can be obtained from Eq.~(\ref{prob_i}) setting $\tau=0$:
\begin{equation}
\Pi(i|\{t_k\})=\frac{\lambda_i(t_i)}{N \bar{\lambda}(\{t_k\})}
\label{prob_i_approx}
\end{equation}
whereas the distribution of the time until the next event is
\begin{equation}
 \varphi(\tau |\{t_k\})=N \bar{\lambda}(\{t_k\}) e^{-N \bar{\lambda}(\{t_k\}) \tau}. 
\label{Delta_time_approx}
\end{equation}
In the Markovian case, $\lambda_i(t_i)=\lambda_i$ and we recover the classical Gillespie algorithm given in Eq.~(\ref{prob_i_gillespie}). In fact, quite remarkably, the new algorithm works as the original Gillespie algorithm with the difference that now the individual rates depend on the elapsed times of the processes and, therefore, are stochastic processes themselves. We name this algorithm generalized ``non-Markovian Gillespie algorithm'' (nMGA).

\begin{figure}[t]
\centerline{\includegraphics[width=\linewidth]{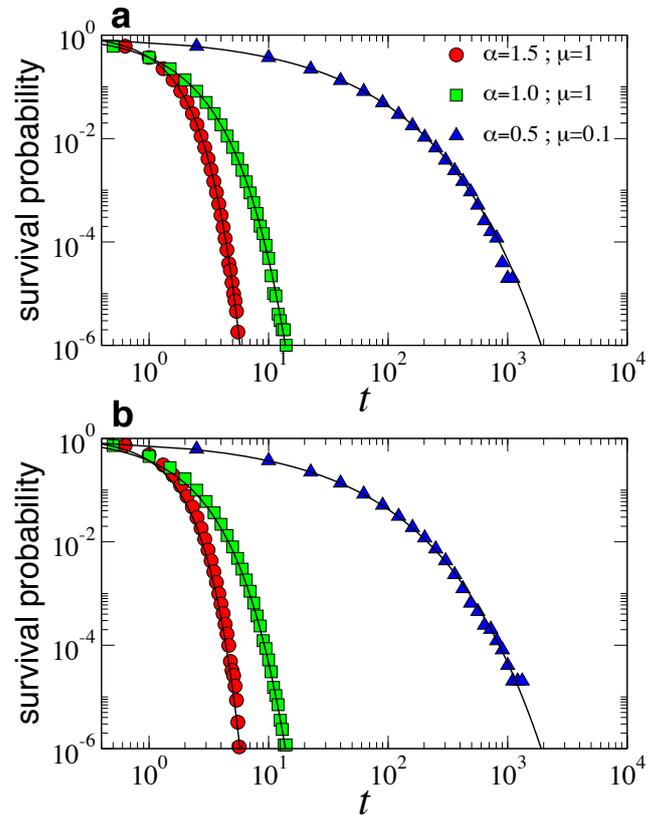}}
\caption{{\bf Testing the algorithm.} Survival probability for three processes generated by the nMGA with parameters $(\alpha=1, \mu=1)$,  $(\alpha=1.5, \mu=1)$, and $(\alpha=0.5, \mu=0.1)$. Solid lines are the given by Eq.~(\ref{weibull}) with the same parameters. In {\bf a} the total number of processes is $N=10^3$ whereas in {\bf b} it is only $N=20$}.\label{fig:1}
\end{figure}

We test the nMGA with a set of $N=10^3$ independent renewal processes. The inter-event time survival probability is taken to be the versatile Weibull distribution 
\begin{equation}
\Psi_i(\tau)=e^{-(\mu_i \tau)^{\alpha_i}}
\label{weibull}
\end{equation}
for all processes. However, the scale and shape parameter of each process, $\mu_i^{-1}$ and $\alpha_i$, are chosen uniformly at random in the intervals $\mu_i \in (0.1,1)$ and $\alpha_i \in (0.5,1.5)$, so that processes with temporal scales that differ in many orders of magnitude are mixed in the simulation. The individual instantaneous rates to be used in Eqs.~(\ref{prob_i_approx}) and (\ref{Delta_time_approx}) are given by 
\begin{equation}
\lambda_i(t_i)=\alpha_i\mu_i^{\alpha_i} t_i^{\alpha_i-1}. 
\label{instantaneous_rate}
\end{equation}
Note that the rates diverge at $t_i=0$ whenever $\alpha_i<1$.
We generate a single long sequence of mixed events according to the nMGA. Then, we measure the inter-event time survival probability for each process. In Fig.~\ref{fig:1}~{\bf a}, we show the comparison between the survival probability for three such processes and the theoretical distribution~Eq.~(\ref{weibull}) with the corresponding parameters. As it can be seen, the agreement is extremely good even when processes with very different time scales are combined. In Fig.~\ref{fig:1}~{\bf b}, we check the effect of having a limited number of processes. Even though the nMGA is only approximate, our numerical simulations indicate that even for a small number of processes (20 in our simulations), the algorithm is able to reproduce the inter-event times very accurately, with a small deviation for processes with $\alpha_i>1$.

Next, we present two relevant examples.
In the first, we shall see the effects of a non-Markovian dynamics and, as opposed to the Markovian case, the importance of the specific details of the laws governing the dynamics. In the second, we compare the computational efficiency of the nMGA with other existing methods.

\section{Epidemic spreading. The SIS model as a case study}

The Susceptible-Infected-Susceptible (SIS) model is one of the simplest and most paradigmatic models of epidemic spreading~\cite{Anderson:1991qr}. In this model, individuals within a contact network can be in two states, either susceptible or infected. Infected individuals remain in this state during a random time and then become susceptible again. Susceptible individuals can become infected if they are in contact with infected neighbors. Except for few exceptions~\cite{Vazquez:2007vn,Min:2011ys,Van-Mieghem:2013db}, epidemic processes are always considered as Markovian so that, in the SIS case, infected individuals recover spontaneously at rate $\beta$ and susceptible ones become infected at rate $\lambda$ times the number of infected neighbors. This dynamics undergoes a phase transition between an absorbing (healthy) phase --where any infectious outbreak disappears exponentially fast-- and an endemic phase with a sustained epidemic activity. This transition takes place at a critical value of the effective infection rate $\lambda_\textrm{eff}=\lambda/\beta$ that depends on the topology of the contact network~\cite{Barrat:2008va,newmanbook,Pastor-Satorras:2001oi,Newman:2002cz,Boguna:2002zl,Boguna:2003gv,Boguna:2003qb}.  Here we consider the SIS dynamics on top of the less structured network, the classical Erdos-Renyi random graph~\cite{Erdos:1959yo}. In this simple model, pairs of nodes out of a set of $N$ nodes are connected with probability $p=\langle k \rangle/N$, where $\langle k \rangle$ is the average degree of the network. In the limit $N \gg 1$, this procedure generates a maximally random graph with a Poisson degree distribution. The critical value for the effective infection rate in these model networks is approximately $\lambda_\textrm{eff}^c=1/\langle k \rangle$~\cite{Kephart:1991}. In the subsequent sections we investigate the role of non-Markovian effects.

\subsection{Independent infections}
In the non-Markovian case, the time that individuals remain infected follows the distribution $\psi_{rec}(\tau)$, in general non-exponential. This means that to apply the nMGA, we have to keep track for each infected individual of the time elapsed since he became infected. The infection process is more involved.  In this subsection, we consider that each active link (connecting a susceptible-infected pair) defines a statistically independent infection process following the distribution $\psi_\textrm{inf}(\tau)$. That is, a susceptible individual connected to a single infected individual will become infected after a random time distributed by $\psi_\textrm{inf}(\tau)$ from the moment the link became active. If the susceptible individual is connected to more than one infected neighbor, each active link is considered as statistically independent so that the infection event will take place at the time of the first firing event of any of the current active links. Because the dynamics is non-Markovian, the infection of a susceptible individual depends not only on the number of active links (infected neighbors) but on the elapsed time of each active link which is, in general, different for each infected neighbor. 

\begin{figure}[t]
\centerline{\includegraphics*[width=\linewidth]{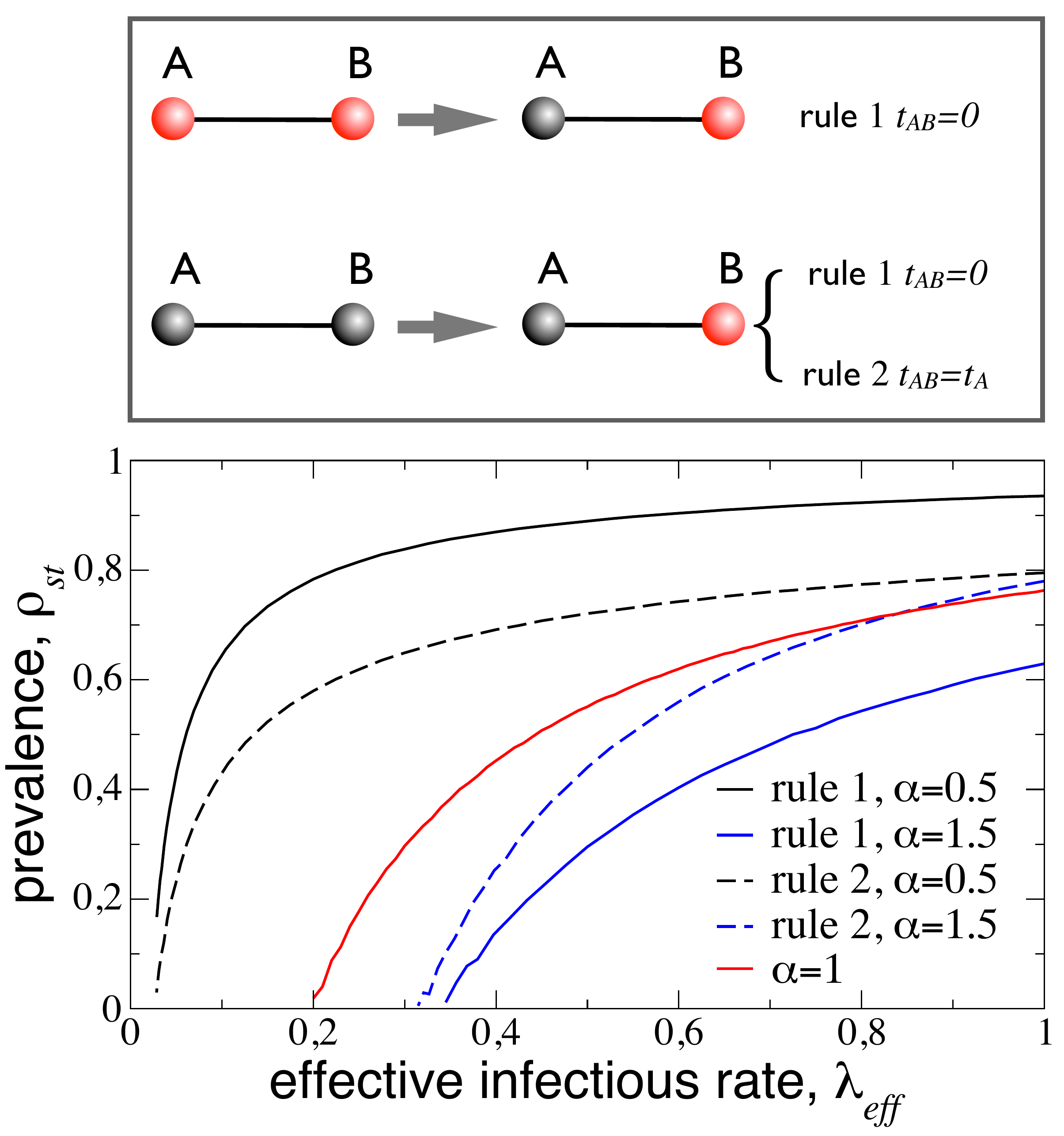}}
\caption{{\bf Non-Markovian epidemic spreading.}{\bf Top:} Two possible ways to generate a new active link in the SIS model. Rules 1 and 2, specify the elapsed time for the newborn active link. Rule 1 sets this time always to zero whereas rule 2 assigns to the newborn active link the elapsed time of individual A. {\bf Bottom:} Prevalence of the epidemics at the steady state as a function of the effective infection rate in Erdos-Renyi networks of size $N=10^4$ and average degree $\langle k \rangle=5$. Solid lines stand for non-Markovian dynamics with the rule 1 implemented whereas dashed lines correspond to rule 2. The red solid curve corresponds to the Markovian case. In all cases, recovery events are exponentially distributed.\label{fig:2}} 
\end{figure}

Therefore, the complexity of the infection course is related to the specific process that leads to the generation of an active link. Indeed, an active link connecting infected individual A and susceptible individual B can reach this configuration from two different
scenarios, as illustrated in the top panel of Fig.~\ref{fig:2}. In the first one, both A and B are originally susceptible and individual A becomes infected by one of his neighbors other than B, generating a new active link. In the second scenario, both A and B are infected and individual B recovers so that an active link is equally created. In the first scenario, it is clear that the active link is new and, therefore, its elapsed time is set to zero, $t_{AB}=0$, we call this ``rule 1''. In the second scenario, we can use again rule 1 and set $t_{AB}=0$. However, we could also argue that infected individual A is the one that makes the action of infection and, thus, we could also consider that the elapsed time of the active link is, in this case, the elapsed time of infected individual A since he became infected, that is, $t_{AB}=t_A$. We call this ``rule 2'' and it is the point of view taken in~\cite{Van-Mieghem:2013db}.     

We first consider the case of Poisson statistics for recovery events and a Weibull distribution with parameter $\alpha$ for the infection process. To compare with the Markovian case, we use as a control parameter a generalization of the effective infection rate, defined as the ratio between the average recovery time and the average infection time, $\lambda_\textrm{eff}=\langle t_{rec} \rangle/\langle t_\textrm{inf} \rangle$. This definition reduces to the effective infection rate used in the Markovian case. Figure~\ref{fig:2} shows the epidemic prevalence (fraction of infected individuals) at the steady state in a network of size $N=10^4$ and average degree $\langle k \rangle=5$ as a function of $\lambda_\textrm{eff}$. As it can be seen in the figure (and also reported in~\cite{Van-Mieghem:2013db}), non-Markovian statistics modifies the position of the critical point significantly. However, there are also important differences between rule 1 and rule 2 for the same values of $\alpha$. For $\alpha>1$, prevalence for rule 2 is always above the one for rule 1 and vice-versa for $\alpha<1$. This difference can be understood by the analysis of the average infection time of an active link, conditioned to a given elapsed time, that is, the first moment of the probability density Eq.~(\ref{eq:2}). In the case of a Weibull distribution, the average time until the next event is an increasing function of the elapsed time when $\alpha<1$, whereas it is decreasing when $\alpha>1$. When a new active link is generated, its elapsed time is always above zero with rule 2 whereas it is exactly zero with rule 1. Therefore, the average infection time with rule 2 is longer or shorter than in the case of rule 1 whenever $\alpha<1$ or $\alpha>1$, respectively.

The effect of a non-Poisson recovery time distribution is much less determinant as compared to the non-Markovian infection dynamics. Indeed, when using rule 1, we do not find any noticeable difference with respect to the Markovian case whereas there are minor differences when using rule 2 but only for very heterogeneous recovery time distributions.

\subsection{Cooperative infections}

One of the consequences of the ``independent infections'' assumption made in the previous subsection is that, for a given individual, the total infection rate at a given time is the sum of the instantaneous rates of all her active links at that time. This is a reasonable assumption when rule 2 is used because, in this case, the infected node is the one associated with the random event of infecting the neighbor whereas the susceptible node is only a passive actor of the process. However, this is not the case when rule 1 is in use because, in such situation, the susceptible node is the one actively associated with the random infectious event. A naive explanation of the difference between these two cases is as follows. For rule 2, we could imagine the infected node firing imprecisely infective agents to her neighbor such that the susceptible node would only become infected after one of these agents hits her. The random infection time is then given by the random time the infected node takes to hit her neighbor, a process attributed solely to the infected node. For rule 1, we could imagine the infected node firing with perfect precision to her neighbor, who is endowed with a protective shield that is destroyed after some exposure to the infective agent and regenerated once the individual becomes susceptible again. However, in this case there is no reason, a priori, to assume the hypothesis of independence between different active links that could act cooperatively and non-linearly to infect the susceptible individual. 

To explore this possibility, we consider a simple example where the total infection rate of a given susceptible node $i$ at time $t$ is
\begin{equation}
\lambda_{tot,i}(t)=\left[ \sum_{j} a_{ij} n_j(t) \left[ \lambda_{i}(t_{ij})\right]^{\frac{1}{\sigma}} \right]^{\sigma},
\label{eq:nonlineal}
\end{equation}
where $a_{ij}$ is the adjacency matrix, $n_j(t)=1$ if node $j$ is infected at time $t$ and zero otherwise, and $t_{ij}$ is the time the link $i-j$ has been active. As before, the instantaneous rate is given by Eq.~(\ref{instantaneous_rate}). For $\sigma=1$ we recover the case of independent infections.
\begin{figure}[t]
\centerline{\includegraphics*[width=\linewidth]{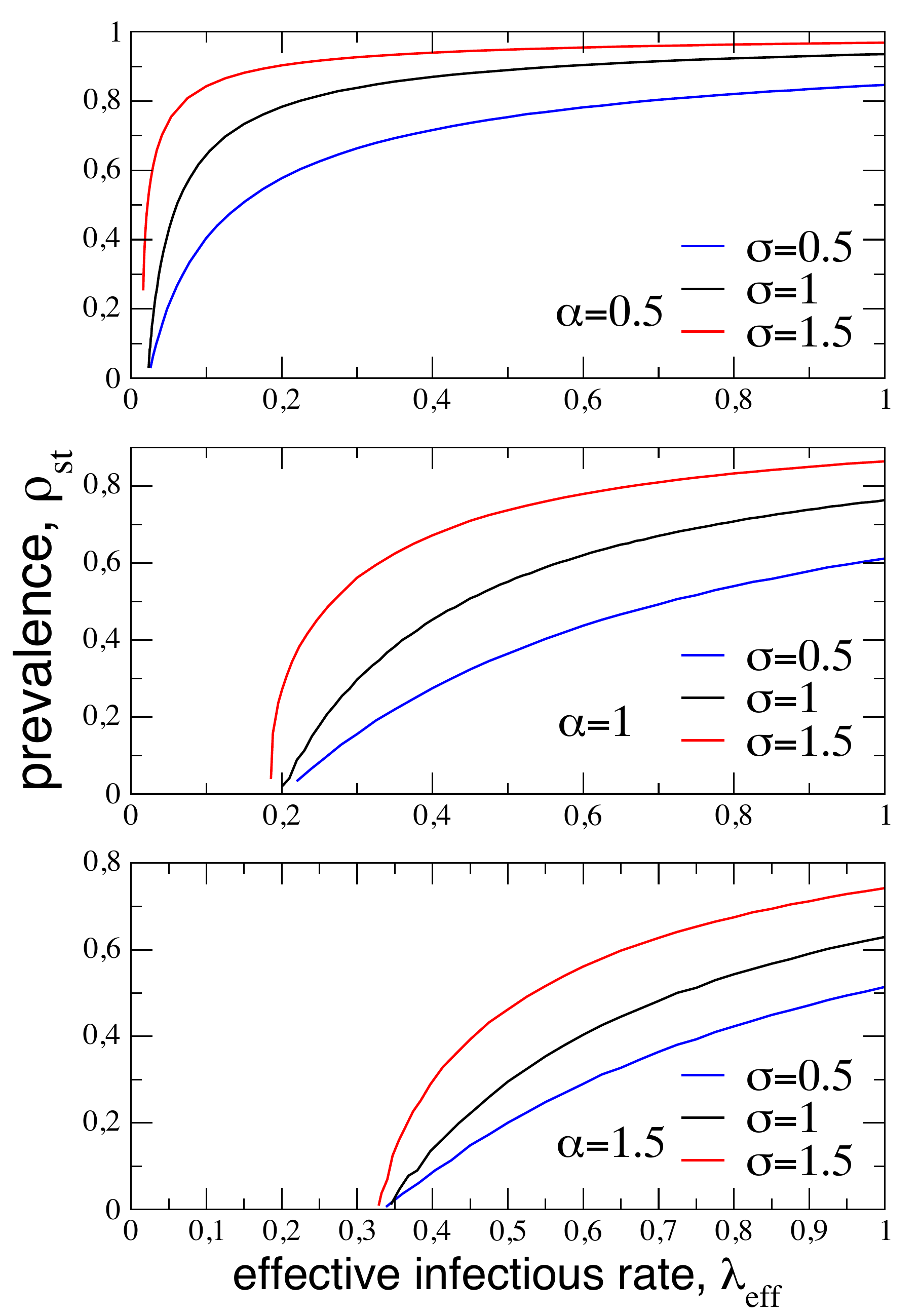}}
\caption{{\bf Cooperative non-Markovian epidemic spreading.} Prevalence of the epidemics at the steady state as a function of the effective infection rate for the same network as in Fig.~\ref{fig:2} in the case of cooperative infections given by Eq.~(\ref{eq:nonlineal}) for different values of $\alpha$ and $\sigma$. Rule 1 is used in all cases.  \label{fig:3}} 
\end{figure}

This case can be readily implemented with the nMGA. Results are shown in Fig.~\ref{fig:3}. Non-linear infections have an important effect on the prevalence of the infection, increasing it when $\sigma>1$ and decreasing it if $\sigma<1$. In this case, however, the position of the critical point is not affected. The reason is that in the low prevalence regime close to the critical point, the number of infected neighbors is very small and, thus, we are effectively in the same regime than in the case of independent infections. It is also possible to implement more complex non-linear schemes, such as threshold models for which the instantaneous infection rate is zero below a given value. In all these cases the nMGA can be applied.

\section{Assessment of the generalized Gillespie algorithm on a system of biochemical reactions with delays}
To show the generality and assess the performance of the nMGA as compared to other existing methods, we apply our approach to a stochastic system of reactions with time delays. Time delays account for the non-Markovian nature of many random processes that play a key role in a wealth of problems in molecular biology involving biochemical reactions or transport. For instance, time delays can model slow processes compound of sequential multistage reactions that can induce stochastic oscillations in gene expression~\cite{Bratsun:2005fk,Miekisz:2011,Lafuerza:2011}. In neurotransmission, time delays can be related to the trap of particles in dendritic spines explaining their anomalous diffusion~\cite{Fedotov:2008kx}. 

This relevance prompted several attempts to adapt Gillespie's algorithm to implement biochemical reactions with time delays for the analysis of gene regulation~\cite{Bratsun:2005fk,Barrio:2006uq,Roussel:2006sf,Cai:2007th,Anderson:2007ad}. When time delays are independent of the evolution of the system the proposed nMGA does not necessarily outperform those previously proposed methods based on annotated lists of future events~\cite{Bratsun:2005fk,Barrio:2006uq,Cai:2007th} or Anderson's modified next reaction algorithm for systems with delays (algorithm 7 in~\cite{Anderson:2007ad}). However, when there is a coupling between the distribution of the time delays and the state of the system, none of the previously developed methods can be straightforwardly applied. The reason is that those methods assume that delay times must be chosen at the moment of the initiation of each reaction, which is clearly not possible if delays depend on the changing state of the system. In Appendix C, we show how to modify Anderson's algorithm to deal with this more general case although, as we shall see below, it is slower than the nMGA.

\begin{figure}[t]
\centerline{\includegraphics[width=\linewidth]{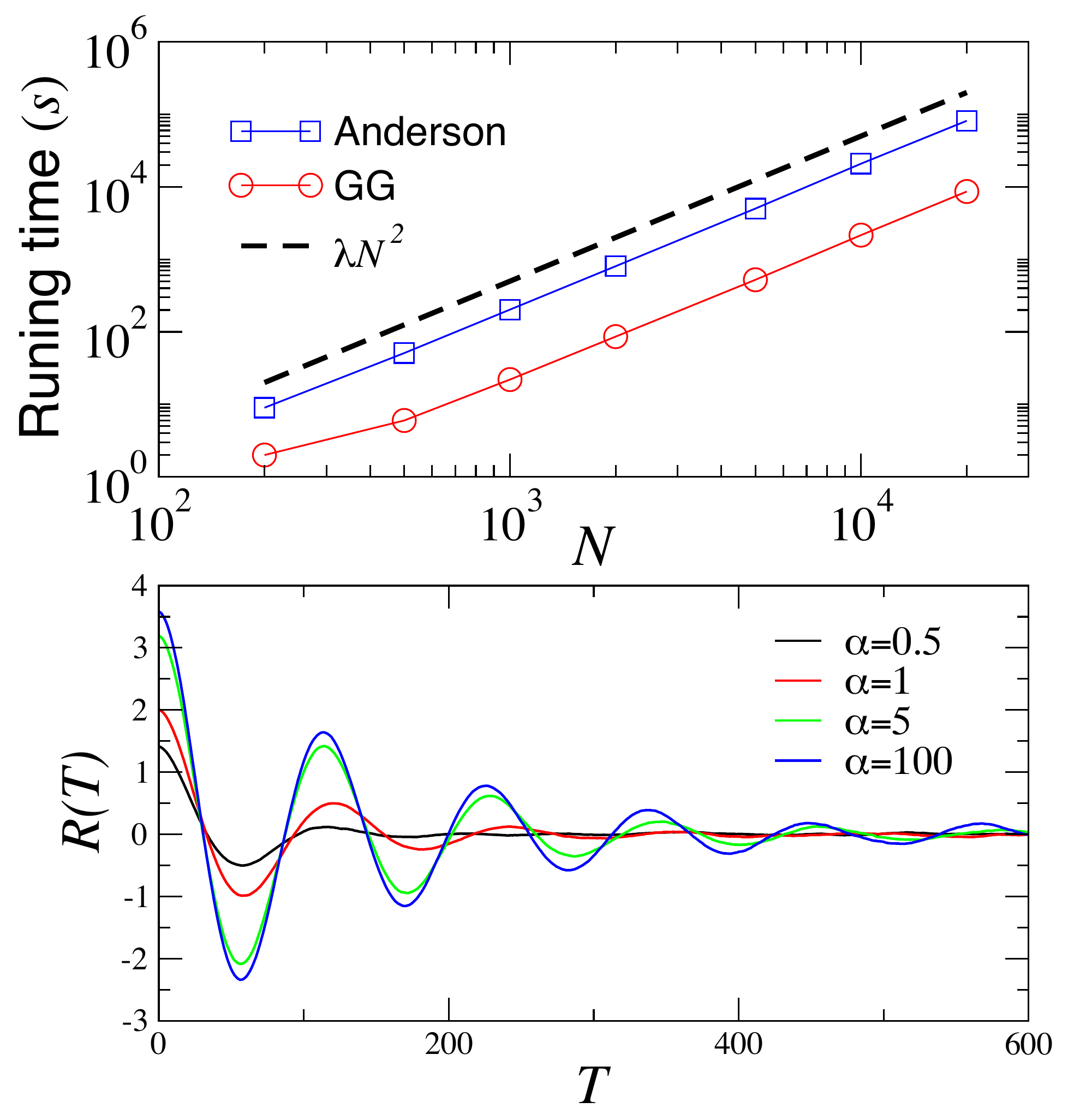}}
\caption{{\bf Performance of the GG and modified Anderson's algorithms for the gene regulation model Eq.~(\ref{schema}) with time delays coupled to the evolution of the system.} (Top) Scaling of computational times required to simulate $10^4$ physical time units, starting with $n_{M*}=0$ and the steady state values of $n_p$ and $n_M$ for and $\alpha=5$. (Bottom) Autocorrelation function, normalized by system size, for the temporal evolution of the number of proteins in the stationary state as a function of the shape parameter $\alpha$ of the time delay distribution, for $N=100$. In both cases the delays follow a Weibull distribution with the parameters detailed in the main text.}
\label{fig:4} 
\end{figure}

As an example, we consider the following stochastic reaction system, which can serve as a model for gene regulation with delayed auto-inhibition
\begin{equation}
\emptyset{{{g(n_P)} \atop \Longrightarrow}\atop{\tau}} M, \hspace{0.4cm}M{{{\beta_P} \atop \longrightarrow}\atop{}} M+P,\hspace{0.4cm}M{{{\gamma_M} \atop \longrightarrow}\atop{}} \emptyset,\hspace{0.4cm}P{{{\gamma_P} \atop \longrightarrow}\atop{}} \emptyset.\label{schema}
\end{equation}  
Here $M$ represents some messenger $RNA$ ($mRNA$) molecule and $P$ is the corresponding translated protein. The generation of $mRNA$ is initiated at a Poissonian process of rate $g(n_P)$ depending on the instantaneous number of proteins $n_P$ present at generation time, but it is completed only after a delay time $\tau$, drawn from a given probability distribution. Translation of $mRNA$ molecules to proteins and spontaneous degradation of $mRNA$ and proteins are modeled by Poissonian processes with rates $\beta_P$, $\gamma_M$, and $\gamma_P$, respectively. A version of this system was analyzed, for example, in~\cite{Jensen:2003,Barrio:2006uq,Miekisz:2011,Lafuerza:2011b,Lafuerza:2011,Lafuerza:2013,Brett:2013kx}. In particular, reference~\cite{Brett:2013kx} considers uniformly distributed delays in the range $(10.7,26.7)$, $\beta_P=1,\gamma_M=\gamma_P=0.03$, and a Hill function $g(n_P)=N\alpha_M/(1+\left(n_P/(NP_0)\right)^h)$ --with $\alpha_M=1, P_0=10, h=4.1$, and variable size $N$-- to reflect that the presence of protein molecules has a negative feedback on $mRNA$ generation. 

As mentioned before, the nMGA is particularly suitable when time delays are coupled to the macroscopic evolution of the system. This is the case, for instance, if we consider a slightly modified gene regulation model where the time required for the transcription of $mRNA$ is affected by the total amount of $mRNA$ being transcribed, $n_{M^*}$, or already present at the time, $n_{M}$, --a likely biological assumption based on the fact that resources needed for $mRNA$ transcription, like nucleotides or ATP energy, are finite.

More specifically, we consider a system-coupled delay model that only differs from the previous one in the distribution of time delays $\tau$. Instead of the uniform distribution, we choose a Weibull distribution with scale parameter $\mu^{-1}=\mu_0^{-1}\left[1+v(n_{M^*}+n_M)/N\right]$ and $\mu_0=0.125$, $v=0.5$. This means that the instantaneous rate of an ongoing process is modified every time the system changes, which implies that the time a reaction takes to complete (the delay) is not defined when the reaction starts. We use this non-Markovian 
version of the model to compare the performance of the nMGA with that of the modified Anderson's algorithm in Appendix C. As shown in Fig.~\ref{fig:4} {\em Top}, the computational time required by both algorithms scales as $N^2$. However, nMGA is a factor $\approx 3.7$ times faster for $\alpha=0.5$ and $\approx 9.5$ times faster for $\alpha=5$. We note, furthermore, that the adapted Anderson's algorithm works relatively well in this case because the cumulative distribution of a Weibull distribution can be expressed in terms of elementary functions. When this is not possible, Anderson's algorithm becomes much slower while the efficiency of nMGA remains the same. In Fig.~\ref{fig:4} {\em Bottom}, we also show the autocorrelation function for the temporal evolution of the number of proteins in the stationary state. As $\alpha$ increases taking values between $0.5$ and $100$, the distribution of time delays becomes markedly more peaked around the mean and, as a consequence, the sequence of oscillations, with decaying amplitude and wave cycle marked by the average delay, becomes more pronounced in the autocorrelation function. Interestingly, as the distribution of time delays drifts away from the exponential becoming more bursty, these oscillations damp out to eventually disappear.

To finish this section, let us comment on a recent and innovative approach proposed in~\cite{Brett:2013kx}, which replaces the master equation by an effective non-linear integro-differential Langevin equation amenable to numerical treatment. The running time of the numerical integration is independent of the system size so, in principle, this method should be preferred over other stochastic methods when the system size is very large (also in this limit the error of the method tends to zero). This will not be the case, however, if one is interested in the highest frequencies of the system, which scale linearly with the system size. Probing such high frequency domain would make it necessary to increase the resolution of the discretization of the integro-differential equation accordingly, with the corresponding increase in computing time. 
In addition, this approach was developed to study well-mixed systems and, thus, it cannot be applied directly to networked populations.

\section{Conclusions}

Models of dynamical processes in complex systems often assume that characteristic random events occur continuously and independently at constant rates. However, this assumption fails for many real systems, which cannot be correctly described unless memory effects like time delays, aging, or bursty dynamics are accounted for using non-Markovian transitions. We have introduced an exact and general framework able to generate statistically correct realizations for systems of non-Markovian discrete stochastic processes. In the limit of a large number of processes, it is approximated to a simple and general simulation algorithm which, quite surprisingly, works exactly as the original Gillespie method with the difference that instantaneous rates of events depend on the time elapsed since the event last took place. Compared with other methods existing in the literature, our algorithm is not always the fastest. However, it is the most general as it can be implemented very easily in cases (like delays coupled to the evolution of the system) where other algorithms do not work or need adapted versions.

Beyond the proven validity and efficiency of the algorithm, our results unveil the drastic effects that very subtle differences in the modeling of non-Markovian processes have on the global behavior of complex systems, with important implications for their understanding and prediction. This turns out to be a central question in many different fields, since evidence shows that non-Markovian dynamics are the rule rather than the exception. For instance, in the Internet information is injected in bursts and packet flow arrival times can experience propagation delays due to congestion and other effects. In human dynamics, bursty behavior affects the way information is generated and spread. In gene regulatory networks, intrinsic stochastic fluctuations may lead to the occurrence of oscillations and other phenomena not observed in Markovian or deterministic analogs. Beyond these examples, the potential range of applications is countless. For all of them, the correct modeling of non-Markovian events is crucial and minimal variations can have drastic effects on their global behavior.

\begin{acknowledgments}
We thank Oleguer Sagarra, Kolja Kleineberg and Pol Colomer-de-Sim\'on for a careful reading of an earlier version of the manuscript and useful comments. We acknowledge support from the James S. McDonnell Foundation 21st Century Science Initiative in Studying Complex Systems Ð Scholar Award; the ICREA Academia prize, funded by the {\it Generalitat de Catalunya}; MINECO projects FIS2010-21781-C02-02,  BFU2010-21847-C02-02 and FIS2012-30634; {\it Generalitat de Catalunya} grant No.\ 2014 SGR 608; and the Ram\'on y Cajal program of the Spanish Ministry of Science; LFL is supported under EPSRC grant EP/H02171X.
\end{acknowledgments}

\appendix

\section{Simulation details for the SIS dynamics}

To apply the nMGA to the SIS model with independent infections, we define two lists. At any given time, the first list contains all infected nodes along with their elapsed times $t_i$ and rates $\lambda_\textrm{rec}(t_i)=\alpha_\textrm{rec}\mu_\textrm{rec}^{\alpha_\textrm{rec}} t_i^{\alpha_\textrm{rec}-1}$, to be used in Eq.~(\ref{prob_i_approx}) in the main text. The second list contains all active links annotated with their elapsed times $t_{ij}$ and rates $\lambda_\textrm{inf}(t_{ij})=\alpha_\textrm{inf}\mu_\textrm{inf}^{\alpha_\textrm{inf}} t_{ij}^{\alpha_\textrm{inf}-1}$. Without loss of generality, we set $\mu_\textrm{rec}=1$ and change $\mu_\textrm{inf}$ to modify the effective infection rate.
We initialize the simulation with all nodes infected so that the list of active links is initially empty. Then, at each step of the simulation we perform the following steps:
\begin{enumerate}
\item
choose next event out of the two lists with the probability given by Eq.~(9) in the main text, ignoring events with elapsed time equal to zero
\item
draw a random time increment $\tau$ from the exponential distribution Eq.~(10) in the main text, $\tau=-\ln(u)/N\overline{\lambda}(\{t_k\})$, with $u$ a uniform random variable in the interval $(0,1)$ and update current time as
\[
t \rightarrow t+\tau
\]
\item
\begin{enumerate}
\item
if the chosen event is the recovery of infected node $i$, 
\begin{enumerate}
\item
remove $i$ from the list of infected nodes
\item
remove also from the list of active links the links between $i$ and his susceptible neighbors
\item
add to the list of active links all the links connecting $i$ with his infected neighbors. These new active links are given elapsed times according to rule 1 or rule 2, depending on the particular choice
\end{enumerate}
\item
if the chosen event is the infection of node $i$ 
\begin{enumerate}
\item
add node $i$ to the list of infected nodes with an elapsed time equal to zero 
\item
remove from the list of active links the links between $i$ and his infected neighbors 
\item
add to the list of active links all the links connecting $i$ with his susceptible neighbors. These new active links are given elapsed times equal to zero
\end{enumerate}
\end{enumerate}
\item
update the elapsed times of the rest of the elements of the lists as
\[
t_i=t_i+\tau \mbox{ and } t_{ij}=t_{ij}+\tau
\]
and go to step 1. In practice, this step can be avoided if we annotate the elements of the lists with their birth-times. Elapsed times can be then evaluated at runtime as current time minus birth time.
\end{enumerate}

In the case of cooperative infections, we generate a new list containing all susceptible nodes annotated with the rates in Eq.~(\ref{eq:nonlineal}). Then, the event in step 1 is chosen from the list of infected nodes and the list of susceptible ones.

\section{Simulations details for delayed reactions}

We give here the details for the reaction system defined by Eq.(\ref{schema}) with delays depending on the state of the system. During the course of the simulation, we keep track of the number of proteins, $n_P$, mRNA's, $n_M$, and a list of the mRNA's that have initiated but not yet completed transcription, $M^*_i$, annotated with their elapsed times since transcription started, $t_i$, and corresponding rates, $\lambda(t_i)=\alpha\mu^{\alpha} t_i^{\alpha-1}$, where $\mu$ depends on the current state of the system as $\mu=\mu_0/(1+v(n_{M^*}+n_M)/N)$. There are five types of events. i) The generation of one protein at rate $n_M\beta_P$, ii) the degradation of one protein at rate $n_P\gamma_P$, iii) the degradation of one mRNA at rate $n_M\gamma_M$, iv) the initiation of transcription of a new mRNA, at a rate $g(n_P)$, and v) the completion of the transcription of a mRNA, at a rate $\alpha\mu^{\alpha} t_i^{\alpha-1}$. Then, at each time step of the simulation:
\begin{enumerate}
 \item
 choose the next event out of these five possibilities according to the probability given by Eq.~(9) in the main text, ignoring mRNA's that just started transcription, i.e., with elapsed time equal to zero. The normalization constant in the denominator of Eq.~(9) is given by $N\overline{\lambda}(\{t_k\})=n_M(\beta_P+\gamma_M)+n_P\gamma_P+g(n_P)+\sum_i \lambda(t_i)$
 \item
 draw a random time increment $\tau$ from the exponential distribution Eq.~(10) in the main text, and update current time as
 \[
 t \rightarrow t+\tau
 \]
 
 \item
 update the elapsed times of all mRNA's in the process of being transcribed in the list
 \[
 t_i=t_i+\tau
 \]
 except for the newly created one, if any, that is set to zero
 \item
 if needed, update the number of molecules $n_P, n_M, n_{M^*}$ and the list of mRNA's in the process of being transcribed and go step 1.
\end{enumerate}

\section{Adaptation of Anderson's modified next reaction method for general non-exponential inter-event times}
Below, we detail how to adapt Anderson's modified next reaction method to general non-exponential inter-event times (including delays coupled to the state of the system). In~\cite{Anderson:2007ad}, it is explained how to use the modified next reaction method (algorithm 3 of~\cite{Anderson:2007ad}) in systems in which the instantaneous rates depend explicitly on time. In the problem we focus on here, the instantaneous rate of a reaction depends on the elapsed time since this reaction last fired. Next, we give a brief explanation of the main ideas behind the adaptation of Anderson's modified next reaction algorithm to treat this case. For details on the original Anderson's method, the reader is referred to~\cite{Anderson:2007ad}. 

As a simple example, we consider a system with a single species. The number of particles of this species at time $t$ is $n(t)$. This number changes by $v_i$ each time reaction $i$ fires, and it does so with an instantaneous rate $r_i(n(t),t-\mathsf{t}_i)$, where $\mathsf{t}_i$ is the last time (before $t$) that reaction $i$ fired. The evolution of $n$ admits a representation using Poisson processes~\cite{Anderson:2007ad}:
\begin{equation}
n(t)=n(0)+\sum_iv_iY_i\left(\int_{0}^tr_i(n(s),s-\mathsf{t}_i)ds\right),\label{Poissonrep}
\end{equation}
where $Y_i$ are independent unit rate Poisson processes (i.e. $P(Y_i(t)=m)=\frac{t^m}{m!}e^{-t}$, $Y_i(t+\Delta)=Y_i(t)+Y(\Delta)$). Note that $Y_i(t)$ is a stochastic process with $Y_i(0)=0$, that increases by one at random times, and that the interval between consecutive increment times are independent random variables exponentially distributed with unit average. Let us now assume that the whole process starts at time $t=0$, so that $\mathsf{t}_i=t=0, \forall i$. The time at which reaction $i$ will next fire, $\mathsf{t}'_i$, is the solution of 
\begin{equation}
\int_{0}^{\mathsf{t}'_i}r_i(n(s),s-\mathsf{t}_i)ds=P_i, \label{nextimeeq}
\end{equation}
with $P_i$ an exponential random variable with unit average (the ``internal time" of  reaction $i$). Next reaction methods \cite{GibsonBruck2000,Anderson:2007ad} are based on the fact that these ``internal times", $P_i$, are random variables independent of the state of the system, and the actual firing times and the subsequent changes in the state of the system can be obtained through (\ref{nextimeeq}) once the $P_i$'s are set.

From a computational perspective (\ref{nextimeeq}) is not yet useful, because it requires the knowledge of $n(t)$ from the current time ($t=0$) until the time of the next firing of reaction $i$, $\mathsf{t}'_i$. However, equation (\ref{nextimeeq}) will be valid for the reaction that  fires first once $n(s)$ is replaced with $n(0)$, since $n(s)=n(0)$ until the first process fires. So one can obtain the time of the next reaction, $t$, by solving (\ref{nextimeeq}) with $n(s)=n(0)$ for each reaction, and then setting $t=\min_i\{\mathsf{t}'_i\}$. One can then update the state of the system ($n$) according to the reaction that fired. The time of the next firing of a reaction $i$ that was not the first firing is given by:
\begin{eqnarray}
&&\int_{0}^{\mathsf{t}'_i}r_i(n(s),s-\mathsf{t}_i)ds=\\
&&\int_{0}^{t}r_i(n(0),s-\mathsf{t}_i)ds+\int_{t}^{\mathsf{t}'_i}r_i(n(t),s-\mathsf{t}_i)ds=P_i,\nonumber
\end{eqnarray}
which again is valid only if reaction $i$ is the one first firing after $t$. 
Defining the ``elapsed internal time", $T_i$, as $T_i\equiv\int_{0}^{t}r_i(n(0),s-\mathsf{t}_i)ds$, we obtain an equation for $\mathsf{t}'_i$ that only uses the value of $n(t)$ at the current time (plus the internal time and the elapsed internal time):
\begin{equation}
\int_{t}^{\mathsf{t}'_i}r_i(n(t),s-\mathsf{t}_i)ds=P_i-T_i. 
\label{eqC4}
\end{equation}
The process can be carried on iteratively. 

The final algorithm, then, proceeds as follows:

\begin{enumerate}
\item
initialize: set $t=0$, $\mathsf{t}_i=0$, $T_i=0$. The state of the system is given by $n$
\item
set the ``internal times" for the firing of next processes: $P_i=\ln(1/u_i)$   with $u_i$ independent random numbers uniformly distributed in the interval $(0,1)$
\item
obtain the tentative ``physical times" for the next firing of the reactions, $\mathsf{t}'_i$, solving Eq.~(\ref{eqC4})
\item
obtain the actual time of the firing of the next reaction, $\Delta = \min_i \{\mathsf{t}'_i\}$; let $\mu$ be the index of the process that actually fires
\item
update the ``elapsed internal times", $T_i =T_i + \int_{t}^{\Delta} r_i(n,s-\mathsf{t}_i) ds$ ;  set a new internal time for the reaction that fired, $P_\mu = P_\mu + \ln(1/u_\mu)$
\item
update time and last firing time of reaction $\mu$, $t= \Delta$; $\mathsf{t}_\mu = t$.
\item
update state of the system ($n$); update the rate functions $r_i=r_i(n)$
\item
go to step 3 or quit.
\end{enumerate}


%

\end{document}